\RequirePackage{fix-cm}
\documentclass[smallextended]{svjour3}       
\smartqed  
\usepackage{graphicx}

\usepackage{amsmath}
\usepackage{amssymb}   
\usepackage{mathtools} 
\usepackage{hyperref}
\hypersetup{colorlinks=true,linkcolor=blue,urlcolor=blue,citecolor=blue}
\usepackage{accents}
\usepackage{tensor}
\usepackage[cal=boondox]{mathalfa}
\usepackage{lipsum}
\usepackage{relsize}
\usepackage{color}
\usepackage{comment}

\usepackage[T1]{fontenc}
\journalname{General Relativity and Gravitation}

\newcommand{\uac}[1]{\underaccent{\tilde}{#1}}

\begin{document}

\title{Straightforward Hamiltonian analysis of $BF$ gravity in $n$ dimensions}



\author{Merced Montesinos\href{https://orcid.org/0000-0002-4936-9170} {\includegraphics[scale=0.05]{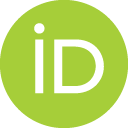}}	\and	Ricardo Escobedo\href{https://orcid.org/0000-0001-5815-4748} {\includegraphics[scale=0.05]{ORCIDiD_icon128x128.png}}	\and Mariano Celada\href{https://orcid.org/0000-0002-3519-4736} {\includegraphics[scale=0.05]{ORCIDiD_icon128x128.png}}}


\institute{Merced Montesinos \at
              Departamento de F\'{i}sica, Cinvestav, Avenida Instituto Polit\'{e}cnico Nacional 2508,\\
              San Pedro Zacatenco, 07360 Gustavo A. Madero, Ciudad de M\'exico, Mexico\\
              \email{merced@fis.cinvestav.mx}           
           \and
           Ricardo Escobedo \at
           Departamento de F\'{i}sica, Cinvestav, Avenida Instituto Polit\'{e}cnico Nacional 2508,\\
           San Pedro Zacatenco, 07360 Gustavo A. Madero, Ciudad de M\'exico, Mexico\\
           \email{rescobedo@fis.cinvestav.mx}
           \and
           Mariano Celada \at
           Centro de Ciencias Matem\'{a}ticas, Universidad Nacional Aut\'{o}noma de M\'{e}xico,\\
           UNAM-Campus Morelia, Apartado Postal 61-3, Morelia, Michoac\'{a}n 58090, Mexico\\
           \email{mcelada@matmor.unam.mx}
}

\date{Received: date / Accepted: date}

\maketitle

\begin{abstract}
We perform, in a manifestly $SO(n-1,1)$ [$SO(n)$] covariant fashion, the Hamiltonian analysis of general relativity in $n$ dimensions written as a constrained $BF$ theory. We solve the constraint on the $B$ field in a way naturally adapted to the foliation of the spacetime that avoids explicitly the introduction of the vielbein. This leads to a form of the action involving a presymplectic structure, which is reduced by doing a suitable parametrization of the connection and then, after integrating out some auxiliary fields, the Hamiltonian form involving only first-class constraints is obtained.
\end{abstract}

\section{Introduction}

In the Lagrangian framework, algebraic constraints on the configuration variables imposed by certain Lagrange multipliers can be handled in two equivalent ways: The first way is to keep the Lagrange multipliers (and the constraint) in the formalism. The second way is to get an equivalent action principle involving less fields by solving explicitly the algebraic constraint. The second way applied to the Plebanski action~\cite{pleb1977118} leads immediately to the Ashtekar Hamiltonian formulation~\cite{ashtekar1986-57-2244,Ashtekar8709} of general relativity {\it without} passing through the self-dual Palatini action~\cite{Samuel1987,Jacobson198739,Jacobsoncqg5} (in terms of the tetrad and the self-dual connection)~\cite{capo1991841} (see also page 10 of the review on $BF$ gravity~\cite{cqgrevBF}). The key point to arrive at the Ashtekar Hamiltonian formulation and {\it not} to the self-dual Palatini action is to solve the simplicity constraint in a form adapted to the foliation of the spacetime. In such a Hamiltonian formulation the solution of the Plebanski 2-forms in terms of the tetrads is never used; it is not needed. More recently, a similar strategy was used to get the Hamiltonian formulation of $BF$ gravity involving the Immirzi parameter (and the cosmological constant)~\cite{BFNoSCC}. Nevertheless, in the latter case, it is not enough to solve the constraint on the $B$ fields as in the Plebanski action because by doing so it leads to a form of the action involving a presymplectic structure, which needs to be further reduced by doing a suitable parametrization of the connection in terms of the final configuration variable and some auxiliary fields. After integrating out the auxiliary fields the Hamiltonian formulation involving only first-class constraints is achieved~\cite{BFNoSCC}. One relevant aspect of such a Hamiltonian formulation is that the tetrad is {\it never} used. However, such a Hamiltonian formulation is the {\it same} as the one we obtain from the Holst action by doing a suitable parametrization of the tetrad and the connection, and after integrating out the auxiliary fields involved~\cite{PhysRevD.101.084003}. In this paper we show that the strategy of Ref.~\cite{BFNoSCC} can also be applied to general relativity in $n$ dimensions expressed as a constrained $BF$ theory. Consequently, this work extends to higher dimensions the results of Ref.~\cite{BFNoSCC}.

{\it Conventions}. We consider a principal bundle over an $n$-dimensional orientable
spacetime manifold $M$ ($n\geq$ 4) that has the global topology
$\mathbb{R}\times\Sigma$, where $\Sigma$ has no boundary. The structure
group is either $SO(n-1,1)$ or $SO(n)$, depending on the signature
involved. Points on $M$ are labeled with coordinates $x^{\mu}=(t,x^a)$,
where $x^a$ ($a,b,\ldots=1,\ldots,n-1$) are adapted coordinates on
$\Sigma$. We use greek letters $\mu,\nu,\ldots=\left\{t,a\right\}$ to
denote spacetime indices, where $t$ refers to the time component. Internal
indices $I,J,\ldots$ take the values $0,\ldots,n-1$ and are raised and
lowered with the internal metric
$(\eta_{IJ})=\text{diag}(\sigma,\underbrace{1,\ldots,1}_{n-1})$, where
$\sigma=-1$ for $SO(n-1,1)$ and $\sigma=1$ for $SO(n)$.  For any kind of
indices, we define sets of $n-4$ and $n-5$ totally antisymmetric indices 
by $[A]$ and $\langle A \rangle$, respectively. The antisymmetrizer is
defined by
\begin{eqnarray}
V^{[A_1 \ldots A_k]}:=\frac{1}{k!}\sum_{P\in S_k}\text{sgn}
(P)V^{A_{P(1)} \ldots A_{P(k)}}.
\end{eqnarray}
The $SO(n-1,1)$ or $SO(n)$ totally antisymmetric tensor $\epsilon_{I_1
	\ldots  I_n}$ satisfies $\epsilon_{01\ldots n-1}=1$. Similarly, the
totally antisymmetric tensor density of weight $+1$ ($-1$) is denoted by
$\tilde{\eta}^{\mu_{1} \ldots \mu_{n}}$
($\underaccent{\tilde}{\eta}_{\mu_{1} \ldots \mu_{n}}$) and fulfills
$\tilde{\eta}^{t 1 \ldots n-1}=1$ ($\underaccent{\tilde}{\eta}_{t 1 \ldots
	n-1}=1$).

\section{First action principle}

General relativity in $n$ dimensions can be written as a constrained $BF$ theory~\cite{Freid_Puz}. There are two action principles to do it. Both action principles include a term with a product of $B$'s that can be antisymetrized in either the spacetime or the internal indices. We are going to study both action principles and show that they both lead to the same canonical theory. Let us first consider the $BF$-type action that involves the antisymetrization in the spacetime indices of the product of $B$'s, which is given by
	\begin{eqnarray}
	S[A,\tilde{B},\Phi,\tilde{\mu}]=&&\int_M d^nx\left(\tilde{B}^{\mu\nu IJ}F_{\mu\nu IJ}+\Phi^{[\alpha]IJKL}\uac{\eta}_{[\alpha]\mu\nu\lambda\rho}\tilde{B}^{\mu\nu}{}_{IJ}\tilde{B}^{\lambda\rho}{}_{KL}\right.\notag\\
	&&\left.+\tilde{\mu}_{[\alpha]}{}^{[M]}\epsilon_{[M]IJKL}\Phi^{[\alpha]IJKL}\right),\label{BFact1}
	\end{eqnarray}
where $\tilde{B}^{\mu\nu}{}_{IJ}$ are bivectors of weight  $1$ taking values in the algebra $\mathfrak{so}(n-1,1)$ or $\mathfrak{so}(n)$; $F_{\mu\nu}{}^{IJ}:=\partial_{\mu} A_{\nu}{}^{IJ} - \partial_{\nu} A_{\mu}{}^{IJ} + A_{\mu}{}^I{}_K A_{\nu}{}^{KJ} - A_{\nu}{}^I{}_K A_{\mu}{}^{KJ}$ is the curvature of the $SO(n-1,1)$ or $SO(n)$ connection $A_{\mu IJ}$; $\Phi^{[\alpha]IJKL}$ is a tensor in both spacetime and internal indices that  satisfies $\Phi^{[\alpha]IJKL}=-\Phi^{[\alpha]JIKL}=-\Phi^{[\alpha]IJLK}=\Phi^{[\alpha]KLIJ}$; and $\tilde{\mu}_{[\alpha]}{}^{[M]}$ is a tensor density of weight $1$. Both $\Phi^{[\alpha]IJKL}$ and $\tilde{\mu}_{[\alpha]}{}^{[M]}$ play the role of Lagrange multipliers. An important fact about this action is that it does not feature a spacetime metric at all, which manifests the background independence of the theory; the metric itself can be regarded as a derived object. The action~\eqref{BFact1} is not the one extensively used in Ref.~\cite{Freid_Puz}, but it can be roughly obtained from it by exchanging the roles of the spacetime and internal indices in the last two terms. Notice that the analog of the last term of~\eqref{BFact1}, which involves $\tilde{\mu}_{[\alpha]}{}^{[M]}$ and imposes an additional condition on $\Phi^{[\alpha]IJKL}$, is not explicitly exhibited in the action of Ref.~\cite{Freid_Puz}. 

Splitting the objects involved in the action~\eqref{BFact1} into their temporal and spatial components, we obtain
\begin{eqnarray}
S=&&\int_{\mathbb{R}} dt\int_{\Sigma}d^{n-1}x\Bigl[ \tilde{\Pi}^{aIJ}F_{taIJ}+\tilde{B}^{abIJ}F_{abIJ}\notag\\
&&+\Phi^{[d]IJKL}\left(2\uac{\eta}_{tabc[d]}\tilde{\Pi}^a{}_{IJ}\tilde{B}^{bc}{}_{KL}+\tilde{\mu}_{[d]}{}^{[M]}\epsilon_{[M]IJKL}\right)\notag\\
&&+(n-4)\Phi^{t\langle e\rangle IJKL}\left(\uac{\eta}_{t\langle e\rangle abcd}\tilde{B}^{ab}{}_{IJ}\tilde{B}^{cd}{}_{KL}+\tilde{\mu}_{t\langle e\rangle}{}^{[M]}\epsilon_{[M]IJKL}\right)\Bigr],\label{BFact2}
\end{eqnarray}
where $\tilde{\Pi}^{aIJ}:=2\tilde{B}^{taIJ}$. Notice that the last line of~\eqref{BFact2} is only present for $n\geq 5$ since the whole term vanishes for $n=4$. The equations of motion for $\Phi^{[d]IJKL}$ and $\Phi^{t\langle e\rangle IJKL}$ are given by
\begin{subequations}
	\begin{eqnarray}
	&&\uac{\eta}_{tabc[d]}(\tilde{\Pi}^a{}_{IJ}\tilde{B}^{bc}{}_{KL}+\tilde{\Pi}^a{}_{KL}\tilde{B}^{bc}{}_{IJ})+\tilde{\mu}_{[d]}{}^{[M]}\epsilon_{[M]IJKL}=0,\label{constr1}\\
	&&\uac{\eta}_{t\langle e\rangle abcd}\tilde{B}^{ab}{}_{IJ}\tilde{B}^{cd}{}_{KL}+\tilde{\mu}_{t\langle e\rangle}{}^{[M]}\epsilon_{[M]IJKL}=0; \label{constr2}
	\end{eqnarray}
\end{subequations}
respectively. Our task is to solve these equations for $\tilde{\Pi}^a{}_{IJ}$, $\tilde{B}^{ab}{}_{IJ}$, $\tilde{\mu}_{[d]}{}^{[M]}$, and $\tilde{\mu}_{t\langle e\rangle}{}^{[M]}$, although the expressions of the last two are fixed once the expressions for the first two are known. Notice that~\eqref{constr1} is linear in 
$\tilde{\Pi}^a{}_{IJ}$ and, independently, also linear in $\tilde{B}^{ab}{}_{IJ}$; whereas~\eqref{constr2} only depends on $\tilde{B}^{ab}{}_{IJ}$ in a quadratic fashion. As mentioned above, equation~\eqref{constr2} only exists for $n\geq 5$, so, for $n=4$ we just have to deal with~\eqref{constr1}. It turns out that, in order to solve both equations~\eqref{constr1} and~\eqref{constr2}, it is enough to consider just~\eqref{constr1}, which means that the set of equations~\eqref{constr2} gives rise to reducibility conditions for the whole system of equations (for $n>4$, of course).

Multiplying~\eqref{constr1} and~\eqref{constr2} by $\epsilon^{[N]IJKL}$, we obtain
\begin{subequations}
	\begin{eqnarray}
	&&\tilde{\mu}_{[d]}{}^{[M]}=-\frac{2\sigma}{4!(n-4)!}\uac{\eta}_{tabc[d]}\epsilon^{IJKL[M]}\tilde{\Pi}^a{}_{IJ}\tilde{B}^{bc}{}_{KL},\label{mu1}\\
	&&\tilde{\mu}_{t\langle e\rangle}{}^{[M]}=-\frac{\sigma}{4!(n-4)!}\uac{\eta}_{tabcd\langle e\rangle}\epsilon^{IJKL[M]}\tilde{B}^{ab}{}_{IJ}\tilde{B}^{cd}{}_{KL},\label{mu2}
	\end{eqnarray}
\end{subequations}
which express both $\tilde{\mu}_{[d]}{}^{[M]}$ and $\tilde{\mu}_{t\langle e\rangle}{}^{[M]}$ in terms of $\tilde{\Pi}^a{}_{IJ}$ and $\tilde{B}^{ab}{}_{IJ}$. These expressions are then substituted back into~\eqref{constr1} and~\eqref{constr2}. Note that the resulting expression from~\eqref{constr2} does not involve $\tilde{\Pi}^a{}_{IJ}$.

In what follows we obtain the solution for $\tilde{\Pi}^a{}_{IJ}$ and $\tilde{B}^{ab}{}_{IJ}$ using only~\eqref{constr1}. This is a remarkable fact because~\eqref{constr1} and~\eqref{constr2} define a coupled system for these variables. Alternatively, the solution of~\eqref{constr1} and~\eqref{constr2} can also be obtained from solving  first~\eqref{constr2} and then~\eqref{constr1}. So, let us start just with~\eqref{constr1}. It can be equally rewritten as
\begin{equation}
\tilde{\Pi}^{[a}{}_{IJ}\tilde{B}^{bc]}{}_{KL}+\tilde{\Pi}^{[a}{}_{KL}\tilde{B}^{bc]}{}_{IJ}+\tilde{\tilde{\mathcal{V}}}{}^{abc}{}_{IJKL}=0,\label{constr3}
\end{equation}
where we have defined
\begin{equation}
\tilde{\tilde{\mathcal{V}}}{}^{abc}{}_{IJKL}:=\frac{1}{3!(n-4)!}\tilde{\eta}^{tabc[d]}\tilde{\mu}_{[d]}{}^{[M]}\epsilon_{[M]IJKL}, \label{V}
\end{equation}
with $\tilde{\mu}_{[d]}{}^{[M]}$ given by~\eqref{mu1}. As a consequence of this definition, $\tilde{\tilde{\mathcal{V}}}{}^{abc}{}_{IJKL}$ is totally antisymmetric in both spacetime and internal indices separately.

The solution of the constraint~\eqref{constr3} involves just $n^2$ fields $\tilde{\Pi}^{a}{}_{I}$, $\uac{N}$, and $N^a$. It is given by (see details in Ref.~\cite{BFn_details})
\begin{subequations}
	\begin{eqnarray}
	&&\tilde{\Pi}^{a}{}_{IJ} =-2\tilde{\Pi}^a{}_{[I} n_{J]}, \label{SolPi2}\\
	&&\tilde{B}^{ab}{}_{IJ} = N^{[a}(\tilde{\Pi}^{b]}{}_{I} n_{J}-\tilde{\Pi}^{b]}{}_{J} n_{I})+\sigma \uac{N} \tilde{\Pi}^{[a}{}_{I}\tilde{\Pi}^{b]}{}_{J}, \label{BIJ6}
	\end{eqnarray}
\end{subequations}
with $n_I$ completely determined by $\tilde{\Pi}^{a}{}_{I}$ as
\begin{equation}
\label{mPi}
n_{I}=\frac{1}{(n-1)! \sqrt{h}}\epsilon_{IJ_1 \ldots J_{n-1}}\underaccent{\tilde}{\eta}_{t a_1 \ldots a_{n-1}} \tilde{\Pi}^{a_1 J_1}  \cdots \tilde{\Pi}^{a_{n-1} J_{n-1}},
\end{equation}
where $h:=\mathrm{det}(\tilde{\tilde{h}}{}^{ab})$, for $\tilde{\tilde{h}}{}^{ab}:=\tilde{\Pi}^{a I}\tilde{\Pi}^{b}{}_{I}$, is positive definite and has weight $2(n-2)$. This latter object allows us to bring in the spatial metric on $\Sigma$, which can be defined as $q_{ab}:=h^{\frac{1}{n-2}}\uac{\uac{h}}{}_{ab}$ with $\uac{\uac{h}}{}_{ab}$ the inverse of $\tilde{\tilde{h}}{}^{ab}$~\cite{PalatininD}. Notice that~\eqref{constr3} is linear and homogeneous in both $\tilde{\Pi}^{a}{}_{IJ}$ and $\tilde{B}^{ab}{}_{IJ}$. Therefore, if $(\tilde{\Pi}^{a}{}_{IJ},\tilde{B}^{ab}{}_{IJ})$ solves~\eqref{constr3}, so does $(\pm\tilde{\Pi}^{a}{}_{IJ},\pm\tilde{B}^{ab}{}_{IJ})$, where both signs are independent from one another. Nevertheless, the sign in front of $\tilde{B}^{ab}{}_{IJ}$ can always be eliminated by suitably redefining the signs in front of the variables $\uac{N}$, $N^a$ and $\tilde{\Pi}^{a}{}_{I}$. On the other hand, because in odd-dimensional spacetimes $\tilde{\Pi}^{a}{}_{IJ}$ given by~\eqref{SolPi2} switches sign under the change $\tilde{\Pi}^{a}{}_{I}\rightarrow -\tilde{\Pi}^{a}{}_{I}$, but remains invariant under the same change in even-dimensional spacetimes, the sign in front of $\tilde{\Pi}^{a}{}_{IJ}$ can only be eliminated in odd-dimensional spacetimes. This fact can be observed more directly by noting that the number of factors of $\tilde{\Pi}^{a}{}_{I}$ in $n^{I}$ depends on the dimension of the manifold. Without loss of generality, below we just consider the solution for $(\tilde{\Pi}^{a}{}_{IJ}, \tilde{B}^{ab}{}_{IJ})$ as given by \eqref{SolPi2} and \eqref{BIJ6}; any other case involving a different sign in front of them can be either redefined as explained before or treated likewise. 

We point out that the solution~\eqref{BIJ6} also solves~\eqref{constr2}, fixing at the same time the value of the variables $\tilde{\mu}_{t\langle e\rangle}{}^{[M]}$. Thus, equations~\eqref{constr2} are in some sense redundant and in consequence represent reducibility relations for the whole system of equations \eqref{constr1} and \eqref{constr2}. To quantify the number of reducibility relations involved in these equations (this reducibility is related to an additional gauge symmetry of the Lagrange multipliers $\Phi^{[\alpha]IJKL}$ in~\eqref{BFact1}; see Ref.~\cite{Freid_Puz}), notice that  they represent $\frac{1}{2}\binom{n}{4}\binom{n}{2}\left[ \binom{n}{2}+1 \right]$ initial equations $l$ for the objects $\tilde{\Pi}^{a}{}_{IJ}$, $\tilde{B}^{ab}{}_{IJ}$, $\tilde{\mu}_{[a]}{}^{[I]}$ and $\tilde{\mu}_{t \langle e \rangle}{}^{[I]}$, which in turn amount to $\binom{n}{2}^{2}+\binom{n}{4}^{2}$ independent unknowns $u$. Then, according to~\eqref{SolPi2} and~\eqref{BIJ6}, the solution for $\tilde{\Pi}^{a}{}_{IJ}$ and $\tilde{B}^{ab}{}_{IJ}$ [as well as the solution for the $\mu$'s, see~\eqref{mu1} and~\eqref{mu2}] involves $n^{2}$ free variables $v$ that are given by $\tilde{\Pi}^{a}{}_{I}$, $\uac{N}$  and $N^{a}$. Note that only for the case $n=4$, the number of relations $(l+v)$ coincides with the number of independent unknowns $u$, so that there are no reducibility relations in this case. However, as mentioned above, for $n>4$ the equations defined by \eqref{constr1} and \eqref{constr2} are not independent from each other. To get the number of reducibility relations $r$ we must subtract from the quantity $(l+v)$ the number of independent unknowns $u$, resulting in $r=(l+v)-u=\frac{1}{288}(n-4)(n-3)n^{2}(n+1)(n+2)(n^{2}+2n+9)$. This number vanishes for $n=4$ as stated before. Therefore, the number of truly independent equations in \eqref{constr1} and \eqref{constr2} is given by $l-r=\frac{1}{576}(n-3)n^{2}(n^{5}-9n^{4}+31n^{3}-51n^{2}+184n+132)$.


Substituting~\eqref{SolPi2} and~\eqref{BIJ6} in the action~\eqref{BFact2}, we obtain, after integrating by parts the first term and neglecting boundary terms,
\begin{eqnarray}\label{action_Pi}
S &=&\int_{\mathbb{R}} dt\int_{\Sigma}d^{n-1}x \left ( -2\tilde{\Pi}^{a I} n^{J}\partial_{t}A_{a I J} + A_{t I J}\tilde{\mathcal{G}}^{I J} - N^{a}\tilde{\mathcal{V}}_{a} - \underaccent{\tilde}{N} \tilde{\tilde{\mathcal{C}}} \right ),
\end{eqnarray}
with 
\begin{subequations}
	\begin{eqnarray}
	&&\label{gauss} \tilde{\mathcal{G}}^{IJ}:=-2D_a\bigl(\tilde{\Pi}^{a[I} n^{J]}\bigr),\\
	&&\label{vector} \tilde{\mathcal{V}}_{a}:= -2\tilde{\Pi}^{bI} n^{J} F_{a b I J}, \\
	&&\label{scalar} \tilde{\tilde{\mathcal{C}}} := -\sigma \tilde{\Pi}^{a I}\tilde{\Pi}^{b J}F_{a b I J}.
	\end{eqnarray}
\end{subequations}

Now, following the same approach of Ref.~\cite{PalatininD}, we realize that the term involving $\partial_t A_{aIJ}$ in~\eqref{action_Pi} can be written as
\begin{equation}
\label{kinetic_term}
-2\tilde{\Pi}^{a I}n^{J}\partial_{t} A_{a I J} = 2 \tilde{\Pi}^{a I}\partial_{t} \left ( W_{a}{}^{b}{}_{I J K} A_{b}{}^{J K} \right ),
\end{equation}
with $W_{a}{}^{b}{}_{IJK} = -W_{a}{}^{b}{}_{IKJ}$ given by
\begin{equation}\label{equation_for_W}
W_{a}{}^{b}{}_{IJK}:= -\left( \delta_{a}^{b} \eta_{I [J}n_{K]} + n_{I} \uac{\uac{h}}_{a c}\tilde{\Pi}^{c}{}_{[J}\tilde{\Pi}^{b}{}_{K]} \right).
\end{equation}
The relation~\eqref{kinetic_term} clearly suggests to define the $n(n-1)$ configuration variables 
\begin{equation}\label{q_equation}
{\mathcal Q}_{a I}:=W_{a}{}^{b}{}_{I J K} A_{b}{}^{J K},
\end{equation}
which are canonically conjugate to $\tilde{\Pi}^{a}{}_{I}$. From~\eqref{q_equation}, the solution for $A_{aIJ}$ reads~\cite{PalatininD}
\begin{equation}
\label{solution_omega}
A_{aIJ} = M_{a}{}^{b}{}_{IJK} {\mathcal Q}_{b}{}^{K} +  \tilde{\tilde{N}}_a{}^{bcd}{}_{IJ}\uac{\uac{\lambda}}_{bcd},
\end{equation}
with
\begin{eqnarray}
M_{a}{}^{b}{}_{I J K} &:=& \frac{2 \sigma}{(n-2)}\bigg{[} (n-2)\delta^{b}_{a}n_{[I}\eta_{J] K} +  \uac{\uac{h}}_{a c}\tilde{\Pi}^{c}{}_{[I}\tilde{\Pi}^{b}{}_{J]}n_{K} \bigg{]},\label{equation_M}\\
\tilde{\tilde{N}}_a{}^{bcd}{}_{IJ}&:=&\left(\delta_a^b\delta_e^{[c}\delta_f^{d]}-\frac{2}{n-2}\uac{\uac{h}}_{ae}\tilde{\tilde{h}}^{b[c}\delta_f^{d]}\right)\tilde{\Pi}^{e}{}_{[I}\tilde{\Pi}^{f}{}_{J]}, \label{equation_N}
\end{eqnarray}
where the variables $\uac{\uac{\lambda}}_{abc}$ satisfy $\uac{\uac{\lambda}}_{abc}= - \uac{\uac{\lambda}}_{acb}$ and the traceless condition $\uac{\uac{\lambda}}_{abc} \tilde{\tilde{h}}^{ab}=0$. Substituting \eqref{solution_omega} into the action \eqref{action_Pi} we get
\begin{equation}\label{Palatini_action_substituting_omega}
S= \int_{\mathbb{R}}dt \int_{\Sigma}d^{n-1}x \bigg{(} 2\tilde{\Pi}^{aI}\partial_{t}{\cal{Q}}_{aI}+A_{tIJ}\tilde{\mathcal{G}}^{IJ}-N^{a}\tilde{\mathcal{V}}_{a}-\uac{N}\tilde{\tilde{\mathcal{C}}} \bigg{)},
\end{equation}
where
\begin{subequations}
\begin{eqnarray}
\label{Gauss_Q_cursive} &&\tilde{\mathcal{G}}^{IJ}= 2 \tilde{\Pi}^{a [I} {\cal{Q}}_{a}{}^{J]} + 4 \delta^{I}_{[K}\delta^{J}_{L]} \tilde{\Pi}^{a [K} n^{M]} \Gamma_{a}{}^{L}{}_{M}, \\
\label{vector_Q_cursive} &&\tilde{\mathcal{V}}_{a} = 2 \bigg{(} 2 \tilde{\Pi}^{bI}\partial_{[a} {\cal{Q}}_{b]I} - {\cal{Q}}_{aI} \partial_{b}\tilde{\Pi}^{bI} \bigg{)} + \tilde{\mathcal{G}}_{IJ}\bigg{(} M_{a}{}^{bIJK} {\cal{Q}}_{bK} + \tilde{\tilde{N}}_a{}^{bcdIJ} \uac{\uac{\lambda}}_{bcd} \bigg{)}, \nonumber \\
\\
\label{scalar_Q_cursive}  && \tilde{\tilde{\mathcal{C}}} = - \sigma \tilde{\Pi}^{aI} \tilde{\Pi}^{bJ} R_{abIJ} +2 \tilde{\Pi}^{a[I}\tilde{\Pi}^{|b|J]} \bigg{(} {\cal{Q}}_{aI} {\cal{Q}}_{bJ} + 2 {\cal{Q}}_{aI} \Gamma_{bJK} n^{K} \nonumber \\
&& \hspace{8mm}  + \Gamma_{aIL}\Gamma_{bJK}n^{K}n^{L} \bigg{)} + 2 \tilde{\Pi}^{a I} n^{J} \nabla_{a}\tilde{\mathcal{G}}_{IJ} - \frac{(n-3)}{(n-2)} \sigma n^{I}\tilde{\mathcal{G}}^{J}{}_{K} n^{K}\tilde{\mathcal{G}}_{IJ} \nonumber \\
&& \hspace{8mm} + \sigma \tilde{\tilde{h}}^{db} \tilde{\tilde{h}}^{cf} \tilde{\tilde{h}}^{ea} \bigg{(} \uac{\uac{\lambda}}_{abc} - \uac{\uac{U}}_{abc}{}^{h}{}_{KL}\Gamma_{h}{}^{KL} \bigg{)} \bigg{(} \uac{\uac{\lambda}}_{dfe} - \uac{\uac{U}}_{dfe}{}^{g}{}_{IJ}\Gamma_{g}{}^{IJ} \bigg{)},
\end{eqnarray}
\end{subequations}
with $\Gamma_{aIJ}$ being the $SO(n-1,1)$ or $SO(n)$ connection compatible with $\tilde{\Pi}^{aI}$, i.e., defined by  $\nabla_{a}\tilde{\Pi}^{b I}= \partial_{a}\tilde{\Pi}^{b I} + \Gamma_{a}{}^{I}{}_{J}\tilde{\Pi}^{b J} + \Gamma^{b}{}_{a c}\tilde{\Pi}^{c I} - \Gamma^{c}{}_{a c}\tilde{\Pi}^{b I} =0$ with $\Gamma_{aIJ}= - \Gamma_{aJI}$ and $\Gamma^a{}_{bc}=\Gamma^a{}_{cb}$, and $R_{abIJ}:= \partial_{a}\Gamma_{bIJ} - \partial_{b}\Gamma_{aIJ} + \Gamma_{aIK} \Gamma_b{}^K{}_J - \Gamma_{bIK} \Gamma_a{}^K{}_J$ being its curvature. Then, collecting all the terms involving $\tilde{\mathcal{G}}^{IJ}$ in  $\tilde{\mathcal{V}}_{a}$ and $\tilde{\tilde{\mathcal{C}}}$, and renaming the term multiplying $\tilde{\mathcal{G}}^{IJ}$ by $\Lambda_{IJ}$, the action takes the form
\begin{equation}\label{Palatini_action_factoring_lambda}
S = \int_{\mathbb{R}}dt \int_{\Sigma}d^{n-1}x \bigg{(} 2\tilde{\Pi}^{a I}\partial_{t} {\cal{Q}}_{a I} -\Lambda_{IJ} \tilde{\mathcal{G}}^{IJ} -2N^{a}\tilde{\mathcal{D}}_{a} - \uac{N} \tilde{\tilde{\mathcal{S}}} \bigg{)}, 
\end{equation}
for
\begin{subequations}
\begin{eqnarray}
	\label{Gauss_Q_cursive_2} && \tilde{\mathcal{G}}^{IJ}= 2 \tilde{\Pi}^{a [I} {\cal{Q}}_{a}{}^{J]} + 4 \delta^{I}_{[K}\delta^{J}_{L]} \tilde{\Pi}^{a [K} n^{M]} \Gamma_{a}{}^{L}{}_{M}, \\
	\label{diffeomorphism_Q_cursive_2} && \tilde{\mathcal{D}}_{a}:= 2\tilde{\Pi}^{b I} \partial_{[a} {\cal{Q}}_{b] I} - {\cal{Q}}_{aI}\partial_{b}\tilde{\Pi}^{bI}, \\ 
	\label{scalar_Q_cursive_2} && \tilde{\tilde{\mathcal{S}}}:= -\sigma \tilde{\Pi}^{aI}\tilde{\Pi}^{bJ}R_{abIJ} \nonumber \\
	&& \hspace{8mm} + 2\tilde{\Pi}^{a [I}\tilde{\Pi}^{|b|J]} \bigg{(} {\cal{Q}}_{aI} {\cal{Q}}_{bJ} +2 {\cal{Q}}_{aI} \Gamma_{bJK}n^{K} + \Gamma_{aIK}\Gamma_{bJL} n^{K}n^{L} \bigg{)} \nonumber \\
	&& \hspace{8mm} + \sigma \tilde{\tilde{h}}^{db} \tilde{\tilde{h}}^{cf} \tilde{\tilde{h}}^{ea} \bigg{(} \uac{\uac{ \lambda}}_{abc} - \uac{\uac{U}}_{abc}{}^{h}{}_{KL}\Gamma_{h}{}^{K L} \bigg{)} \bigg{(} \uac{\uac{\lambda}}_{dfe} - \uac{\uac{U}}_{dfe}{}^{g}{}_{I J}\Gamma_{g}{}^{IJ} \bigg{)},	
\end{eqnarray}
\end{subequations}
where $A_{tIJ}$ is related to $\Lambda_{IJ}$ via
\begin{eqnarray}\label{Lambda_redefinition}
A_{tIJ} &:=& -\Lambda_{IJ} + N^a \bigg{(} M_a{}^b{}_ {IJK} {\cal{Q}}_b{}^K +   \tilde{\tilde{N}}_a{}^{bcd}{}_{IJ}\uac{\uac{\lambda}}_{bcd} \bigg{)} \nonumber \\
&& - 2  \tilde{\Pi}^{a}{}_{[I} n_{J]} \nabla_{a} \underaccent{\tilde}{N} - \sigma\frac{(n-3)}{(n-2)} \underaccent{\tilde}{N} n_{[I} \tilde{\mathcal{G}}_{J]K} n^K,
\end{eqnarray}
and we have introduced the object~\cite{PalatininD}
\begin{equation}\label{definition_U}
\uac{\uac{U}}_{abc}{}^{dIJ}:= \bigg{(} \delta^{d}_{a}\uac{\uac{h}}_{e[b}\uac{\uac{h}}_{c]f}-\frac{2}{n-2}\uac{\uac{h}}_{a[b}\uac{\uac{h}}_{c]f}\delta^{d}_{e}  \bigg{)}\tilde{\Pi}^{e[I}\tilde{\Pi}^{|f|J]}. 
\end{equation}
Notice that there are no time derivatives of the variables $\uac{\uac{\lambda}}_{abc}$ in \eqref{Palatini_action_factoring_lambda} and that they appear in quadratic fashion, meaning that they are  auxiliary fields. Thereby, we can fix them using their own equation of motion. Varying \eqref{Palatini_action_factoring_lambda} with respect $\uac{\uac{\lambda}}_{abc}$ we obtain
\begin{equation}\label{variation_of_action_respect_Lambda}
\delta \uac{\uac{\lambda}}_{abc}: \uac{N} \tilde{\tilde{h}}^{d[b}\tilde{\tilde{h}}^{c]e}\tilde{\tilde{h}}^{af}\left( \uac{\uac{\lambda}}_{dfe} - \uac{\uac{U}}_{dfe}{}^{g}{}_{I J}\Gamma_{g}{}^{I J} \right) =0,
\end{equation}
and since we are working in the nondegenerate case, we have $\uac{N}\neq 0$, which yields 
\begin{equation}\label{solution_for_lambda}
\uac{\uac{\lambda}}_{abc}= \uac{\uac{U}}_{abc}{}^{d}{}_{IJ}\Gamma_{d}{}^{IJ}. 
\end{equation}
Substituting the solution of $\uac{\uac{\lambda}}_{abc}$ into \eqref{Palatini_action_factoring_lambda} we obtain
\begin{eqnarray}\label{final_action}
S &=&\int_{\mathbb{R}} dt\int_{\Sigma}d^{n-1}x\left ( 2\tilde{\Pi}^{a I}\partial_{t} {\cal{Q}}_{aI} -\Lambda_{IJ} \tilde{\mathcal{G}}^{I J} -2N^{a}\tilde{\mathcal{D}}_{a} - \underaccent{\tilde}{N} \tilde{\tilde{\mathcal{H}}} \right ),
\end{eqnarray}
where $\tilde{\mathcal{G}}^{I J}$, $\tilde{\mathcal{D}}_{a}$, and $\tilde{\tilde{\mathcal{H}}}$ are, correspondingly, the Gauss, diffeomorphism, and scalar constraints, given by
\begin{subequations}
	\begin{eqnarray}
	&&\label{Gauss2} \tilde{\mathcal{G}}^{IJ} =  2 \tilde{\Pi}^{a [I} \mathcal{Q}_{a}{}^{J]} + 4 \delta^{I}_{[K}\delta^{J}_{L]} \tilde{\Pi}^{a [K} n^{M]} \Gamma_{a}{}^{L}{}_{M}, \\
	&&\label{diffeomorphism} \tilde{\mathcal{D}}_{a} = 2\tilde{\Pi}^{b I} \partial_{[a} \mathcal{Q}_{b] I} - \mathcal{Q}_{aI}\partial_{b}\tilde{\Pi}^{bI}, \\ 
	&&\label{scalar2} \tilde{\tilde{\mathcal{H}}} := - \sigma \tilde{\Pi}^{a I}\tilde{\Pi}^{b J}R_{a b I J} +  2 \tilde{\Pi}^{a [I}\tilde{\Pi}^{|b|J]} \left ( \mathcal{Q}_{aI} \mathcal{Q}_{bJ} 
	\right. \nonumber \\
	&& \left. \hspace{9mm}+2 \mathcal{Q}_{aI} \Gamma_{bJK} n^{K} + \Gamma_{aIK} \Gamma_{bJL} n^{K} n^{L} \right ).
	\end{eqnarray}
\end{subequations}
Thus, the variables $({\cal{Q}}_{aI},\tilde{\Pi}^{aI})$ label the points of the kinematic phase space of the theory and are subject to the first-class constraints $\tilde{\mathcal{G}}^{IJ}$, $\tilde{\mathcal{D}}_{a}$ and $\tilde{\tilde{\mathcal{H}}}$.

Furthermore, we can get a more familiar Hamiltonian formulation than the one given by~\eqref{final_action} by applying the following canonical transformation 
\begin{subequations}
	\begin{eqnarray} 
	\label{transformation_Q} Q_{aI}&=& {\cal{Q}}_{aI}-W_{a}{}^{b}{}_{IJK}\Gamma_{b}{}^{JK}, \\
	\label{transformation_Pi} \tilde{\Pi}^{aI}&=& \tilde{\Pi}^{aI}.
	\end{eqnarray}
\end{subequations}
This transformation is canonical because:
\begin{equation}\label{prove_canonical_transformation}
2 \tilde{\Pi}^{a I}\partial_{t}Q_{aI} = 2 \tilde{\Pi}^{a I}\partial_{t} {\cal{Q}}_{aI} + \partial_{a} \left ( 2 n_{I}\partial_{t}\tilde{\Pi}^{a I} \right ). 
\end{equation}
Thus, using~\eqref{transformation_Q} and~\eqref{transformation_Pi}, the action \eqref{final_action} becomes
\begin{eqnarray}\label{Thiemann_action}
S &= &  \int_{\mathbb{R}}dt \int_{\Sigma}d^{n-1}x \left ( 2\tilde{\Pi}^{a I}\partial_{t}Q_{a I} -\Lambda_{IJ} \tilde{\mathcal{G}}^{I J} - 2 N^{a}\tilde{\mathcal{D}}_{a} - \underaccent{\tilde}{N} \tilde{\tilde{\mathcal{H}}} \right ),
\end{eqnarray}
with
\begin{subequations}
	\begin{eqnarray}
	&&\label{GaussQ} \tilde{\mathcal{G}}^{IJ} = 2 \tilde{\Pi}^{a [I}Q_{a}{}^{J]}, \\
	&&\label{diffeomorphismQ} \tilde{\mathcal{D}}_{a} = 2\tilde{\Pi}^{b I} \partial_{[a} Q_{b]I} - Q_{a}{}^{I}\partial_{b}\tilde{\Pi}^{b}{}_{I}, \\
	&&\label{scalarQ} \tilde{\tilde{\mathcal{H}}} = - \sigma \tilde{\Pi}^{aI} \tilde{\Pi}^{bJ}R_{abIJ} +  2 \tilde{\Pi}^{a [I}\tilde{\Pi}^{|b|J]}Q_{aI}Q_{bJ}.
	\end{eqnarray}
\end{subequations}
Notice that the canonical coordinates $(Q_{aI}, \tilde{\Pi}^{aI})$ are $SO(n-1,1)$ or $SO(n)$ vectors. One relevant aspect of the Hamiltonian formulations~\eqref{final_action} and~\eqref{Thiemann_action} is that they were achieved {\it without} involving the orthonormal frame of 1-forms (vielbein) $e^I$. Moreover, the Hamiltonian formulations~\eqref{final_action} and~\eqref{Thiemann_action} are the {\it same} as the ones we obtain from the $n$-dimensional Palatini action by doing a suitable parametrization of the vielbein and the connection, and after integrating out the auxiliary fields involved~\cite{PalatininD} (see also~\cite{Bodendorfer_2013} for a different approach).

\section{Second action principle}

{The original action of Ref.~\cite{Freid_Puz} has the same structure of~\eqref{BFact1}, but now the antisymetrization in the internal indices of the product of $B$'s is regarded. It is given by
	\begin{eqnarray}
	S[A,\tilde{B},\uac{\Phi},\tilde{\nu}]=&&\int_M d^nx\left(\tilde{B}^{\mu\nu IJ}F_{\mu\nu IJ}+\uac{\Phi}_{\mu\nu\lambda\sigma[M]}\epsilon^{[M]IJKL}\tilde{B}^{\mu\nu}{}_{IJ}\tilde{B}^{\lambda\rho}{}_{KL}\right.\notag\\
	&&\left.+\tilde{\nu}_{[\alpha]}{}^{[M]}\tilde{\eta}^{[\alpha]\mu\nu\lambda\sigma}\uac{\Phi}_{\mu\nu\lambda\sigma[M]}\right),\label{BFact3}
	\end{eqnarray}
where both $\uac{\Phi}_{\mu\nu\lambda\sigma[M]}$ (of weight $-1$) and $\tilde{\nu}_{[\alpha]}{}^{[M]}$ (of weight $1$) are Lagrange multipliers, the former having the following symmetries in the spacetime indices: $\uac{\Phi}_{\mu\nu\lambda\sigma[M]}=-\uac{\Phi}_{\nu\mu\lambda\sigma[M]}=-\uac{\Phi}_{\mu\nu\sigma\lambda[M]}=\uac{\Phi}_{\lambda\sigma\mu\nu[M]}$.

Splitting space and time in~\eqref{BFact3} yields
\begin{eqnarray}
S=&&\!\int_{\mathbb{R}} dt\int_{\Sigma}d^{n-1}x\Bigl[ \tilde{\Pi}^{aIJ}F_{taIJ}+\tilde{B}^{abIJ}F_{abIJ}+\uac{\Phi}_{ta tb[M]}\epsilon^{[M]IJKL}\tilde{\Pi}^a{}_{IJ}\tilde{\Pi}^b{}_{KL}\notag\\
&&+2\uac{\Phi}_{tabc[M]}\left(\epsilon^{[M]IJKL}\tilde{\Pi}^a{}_{IJ}\tilde{B}^{bc}{}_{KL}+2\tilde{\nu}_d{}^{[M]}\tilde{\eta}^{tabc[d]}\right)\notag\\
&&+\uac{\Phi}_{abcd[M]}\left(\epsilon^{[M]IJKL}\tilde{B}^{ab}{}_{IJ}\tilde{B}^{cd}{}_{KL}+(n-4)\tilde{\nu}_{t\langle e\rangle}{}^{[M]}\tilde{\eta}^{tabcd\langle e\rangle}\right)\Bigr],\label{BFact4}
\end{eqnarray}
where $\tilde{\Pi}^{aIJ}=2\tilde{B}^{taIJ}$ as above. Note that in contrast to~\eqref{BFact2}, the fact that the Lagrange multiplier $\uac{\Phi}$ is not totally antisymmetric in the spacetime indices produces one more term in the action (the one involving $\uac{\Phi}_{ta tb[M]}$). The variation of~\eqref{BFact4} with respect to $\uac{\Phi}_{ta tb[M]}$, $\uac{\Phi}_{tabc[M]}$, and $\uac{\Phi}_{abcd[M]}$ gives, correspondingly, the constraints
\begin{subequations}
	\begin{eqnarray}
	&&\epsilon^{[M]IJKL}\tilde{\Pi}^a{}_{IJ}\tilde{\Pi}^b{}_{KL}=0,\label{constr12}\\
	&&\epsilon^{[M]IJKL}\tilde{\Pi}^a{}_{IJ}\tilde{B}^{bc}{}_{KL}+2\tilde{\nu}_{[d]}{}^{[M]}\tilde{\eta}^{tabc[d]}=0,\label{constr22}\\
	&& \epsilon^{[M]IJKL}\tilde{B}^{ab}{}_{IJ}\tilde{B}^{cd}{}_{KL}+(n-4)\tilde{\nu}_{t\langle e\rangle}{}^{[M]}\tilde{\eta}^{tabcd\langle e\rangle}=0.\label{constr32}
	\end{eqnarray}
\end{subequations}
In this case, whereas~\eqref{constr12} is independent of $\tilde{B}^{abIJ}$ and~\eqref{constr32} is independent of $\tilde{\Pi}^{aIJ}$, the constraint~\eqref{constr22} blends both components and thereby fixes the relationship between them. Remarkably, the constraint~\eqref{constr12} does not involve the Lagrange multipliers $\tilde{\nu}$'s.

Handling~\eqref{constr22}, we get 
\begin{equation}\label{nu1}
\tilde{\nu}_{[d]}{}^{[M]}=-\frac{2}{4!(n-4)!}\uac{\eta}_{tabc[d]}\epsilon^{IJKL[M]}\tilde{\Pi}^a{}_{IJ}\tilde{B}^{bc}{}_{KL}.
\end{equation}
Similarly, handling~\eqref{constr32}, we obtain
\begin{equation}
\tilde{\nu}_{t\langle e\rangle}{}^{[M]}=-\frac{1}{4!(n-4)!}\uac{\eta}_{tabcd\langle e\rangle}\epsilon^{IJKL[M]}\tilde{B}^{ab}{}_{IJ}\tilde{B}^{cd}{}_{KL}.\label{nu2}
\end{equation}
As the reader can check, the solution of~\eqref{constr12},~\eqref{constr22}, and~\eqref{constr32} with the Lagrange multipliers $\tilde{\nu}$'s given by~\eqref{nu1} and~\eqref{nu2} is the same as the one given by~\eqref{SolPi2} and~\eqref{BIJ6}. Substituting them into the action~\eqref{BFact4} leads to the same intermediate action~\eqref{action_Pi}, which following again the approach of Ref.~\cite{PalatininD} produces the canonical formulation embodied in~\eqref{final_action} and thus to the Hamiltonian formulation~\eqref{Thiemann_action}.

\section{Concluding remarks}
i) We emphasize that our approach begins by solving the constraint on the $B$ field
(simplicity constraint), which leads to an intermediate action principle
that is subsequently reduced first by doing a suitable parametrization of the
spatial $SO(n-1,1)$ [$SO(n)$] connection $A_{aIJ}$, and later by integrating out
the auxiliary fields $\uac{\uac{\lambda}}_{abc}$. Therefore, our approach is an {\it alternative} to the strict 
Dirac's method~\cite{dirac1964lectures}, which requires from the very beginning to
enlarge the original set of variables and to introduce a reducible set of second-class
constraints (see, for instance, Ref.~\cite{Thiemann_2013}). We 
remind the reader that second-class constraints and reducibility issues
are lacking in our approach, although the final action involving
only first-class constraints could also be obtained following Dirac's
classical analysis. More precisely, even though the approach of this paper
and Dirac's classical analysis share the same initial point (the same
action principle) and share also the same final result (an action
principle involving only first-class constraints), these approaches are conceptually very different from each other (on this, see
also Ref.~\cite{Gotay}). Thus, our procedure is more economic, clean and straightforward than Dirac's approach.

ii) We stress that this manifestly $SO(n-1,1)$ [$SO(n)$] covariant Hamiltonian formulation derived {\it for the first time} from the $BF$-type actions~\eqref{BFact1} and from~\eqref{BFact3} involving just first-class constraints, was also achieved without explicitly using the orthonormal frame of $1$-forms (vielbein) $e^I$ in the formalism. Therefore, we have extended to higher dimensions the results of Ref.~\cite{BFNoSCC}.

iii) We have shown that the $BF$-type action principles~\eqref{BFact1} and~\eqref{BFact3} have the {\it same} Hamiltonian formulation (involving only first-class constraints) that the $n$-dimensional Palatini action has~\cite{Bodendorfer_2013,PalatininD}. This needs to be so, and not otherwise, because all three actions describe Einstein's general relativity.


\begin{acknowledgements}
We thank Diego Gonzalez for useful comments and suggestions on a preliminary version of this manuscript. This work was partially supported by Fondo SEP-Cinvestav and by Consejo Nacional de Ciencia y Tecnolog\'{i}a (CONACyT), M\'{e}xico, Grant No.~A1-S-7701. M.~C. gratefully acknowledges the support of a DGAPA-UNAM postdoctoral fellowship.
\end{acknowledgements}

%
%

\bibliographystyle{spphys}       
\bibliography{references}  

\end{document}